\documentclass[twocolumn]{aastex63}
\usepackage{amsmath}

\newcommand{\fullname}{SDSS~J134441.83+204408.3}
\newcommand{\ratio}{$P_{spin}/P_{orb}$}
\newcommand{\cpd}{$\mathrm{cycles~d}^{-1}$}
\newcommand{\swift}{Swift~J0503.7-2819}
\newcommand{\igr}{IGR~J19552+0044}
\newcommand{\beat}{$\omega - \Omega$}

\shorttitle{A highly magnetized, asynchronous mCV}
\shortauthors{Littlefield et al.}

\submitjournal{ApJL}

\begin{document}

	\title{\fullname: A highly asynchronous, short-period magnetic cataclysmic variable with a 56 MG field strength}
	
	\author[0000-0001-7746-5795]{Colin Littlefield}
	\affiliation{Bay Area Environmental Research Institute, Moffett Field, CA 94035 USA}
	
	\author{Paul A. Mason}
	\affiliation{New Mexico State University, MSC 3DA, Las Cruces, NM, 88003, USA}
	\affiliation{Picture Rocks Observatory, 1025 S. Solano Dr. Suite D., Las Cruces, NM 88001, USA}
	
	\author{Peter Garnavich}
	\affiliation{Department of Physics and Astronomy, University of Notre Dame, Notre Dame, IN 46556, USA}
	
	\author[0000-0003-4373-7777]{Paula Szkody}
	\affiliation{Department of Astronomy, University of Washington, Seattle, WA 98195, USA}

	\author{John Thorstensen}
	\affiliation{Department of Physics and Astronomy, 6127 Wilder Laboratory, Dartmouth College, Hanover, NH 03755-3528, USA}
	
	\author{Simone Scaringi}
	\affiliation{Centre for Extragalactic Astronomy, Department of Physics, Durham University, South Road, Durham, DH1 3LE}
	\author{Krystian I\l{}kiewicz}
	\affiliation{Centre for Extragalactic Astronomy, Department of Physics, Durham University, South Road, Durham, DH1 3LE}
	\affiliation{Astronomical Observatory, University of Warsaw, Al. Ujazdowskie 4, 00-478 Warszawa, Poland}
	\author[0000-0001-6894-6044]{Mark R. Kennedy}
	\affiliation{Department of Physics, University College Cork, Cork, Ireland}
	
	\author{Natalie Wells}
	\affiliation{New Mexico State University, MSC 3DA, Las Cruces, NM, 88003, USA}
	\affiliation{Picture Rocks Observatory, 1025 S. Solano Dr. Suite D., Las Cruces, NM 88001, USA}
	
	\correspondingauthor{Colin Littlefield}
	\email{clittlef@alumni.nd.edu}

	\begin{abstract}

		When the accreting white dwarf in a magnetic cataclysmic variable star (mCV) has a field strength in excess of 10~MG, it is expected to synchronize its rotational frequency to the binary orbit frequency, particularly at small binary separations, due to the steep radial dependence of the magnetic field. We report the discovery of an mCV (\fullname; hereafter, J1344) that defies this expectation by displaying asynchronous rotation (\ratio=0.893) in spite of a high surface field strength (B=56~MG) and a short orbital period (114~min). Previously misidentified as a synchronously rotating mCV, J1344 was observed by TESS during sector 50, and the resulting power spectrum shows distinct spin and orbital frequencies, along with various sidebands and harmonics. Although there are several other asynchronous mCVs at short orbital periods, the presence of cyclotron humps in J1344's SDSS spectrum makes it possible to directly measure the field strength in the cyclotron-emitting region; a previously study estimated 65~MG based on its identification of two cyclotron humps, but we revise this to 56$\pm$2~MG based on the detection of a third hump and on our modeling of the cyclotron spectrum. Short-period mCVs with field strengths above 10~MG are normally expected to be synchronous, so the highly asynchronous rotation in J1344 presents an interesting challenge for theoretical studies of spin-period evolution.

	\end{abstract}
	
	\section{Introduction}
	
	\subsection{Asynchronous magnetic cataclysmic variables}
	
	One of the most-studied aspects of magnetic cataclysmic variable stars (mCVs) is the evolution of the spin period of the primary star, an accreting, magnetized white dwarf (WD). WDs with field strengths above $B\sim$10~MG are expected to rotate synchronously with the binary orbit, such that $P_{spin} = P_{orb}$. These systems are known as polars. At lower field strengths, the WD rotates significantly faster than the binary orbit; these systems are known as intermediate polars (IPs).

	\begin{figure*}
		\centering
		\includegraphics[width=\textwidth]{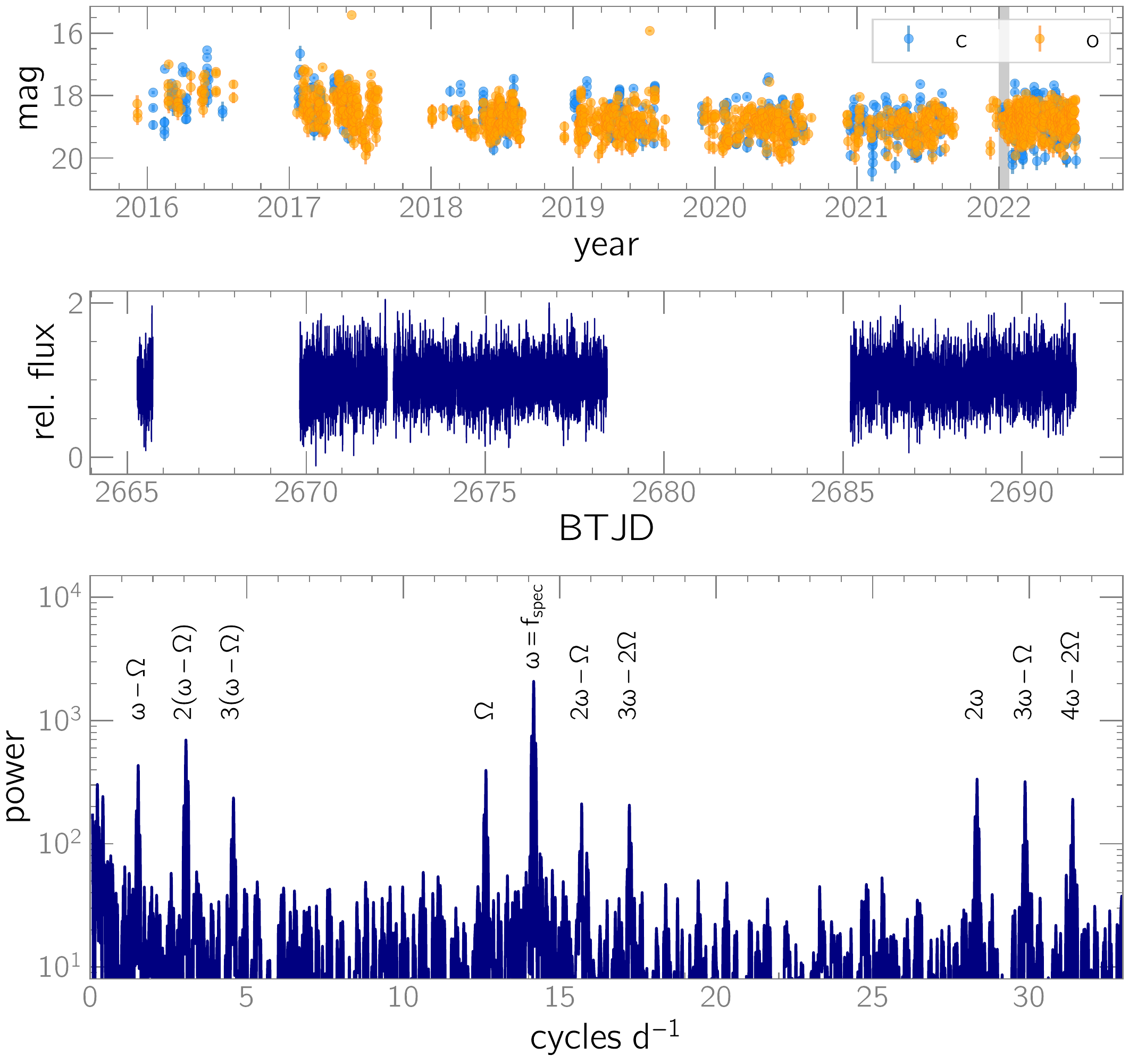}
		\caption{{\bf Top:} Long-term ATLAS light curve of J1344, consisting of c- and o-band observations.  The shaded region indicates the time of the TESS observations. {\bf Middle:} TESS light curve of J1344. {\bf Bottom:} Power spectrum of the TESS light curve. As we discuss in the text, we interpret the \citet{thorstensen20} spectroscopic frequency as the spin frequency $\omega$, but it is conceivable that it is actually the orbital frequency $\Omega$. However, the identification of the beat frequency $\omega-\Omega$ and its harmonics is secure, and the very presence of this frequency establishes that J1344 is an asynchronous rotator. 
		}
		\label{fig:power}
	\end{figure*}

	Somewhat paradoxically, a small number of polars are desynchronized by $\lesssim3\%$. However, these asynchronous polars (APs) are widely thought to be formerly synchronous rotators that are returning to synchronous rotation. A recent nova eruption is the most commonly invoked mechanism for breaking the synchronous rotation. The asynchronous rotation in APs is a temporary disequilibrium, which distinguishes them from IPs, for which there are several proposed equilibrium conditions in which $P_{spin} < P_{orb}$. 
	
	\begin{figure*}
		\centering
		\includegraphics[width=\textwidth]{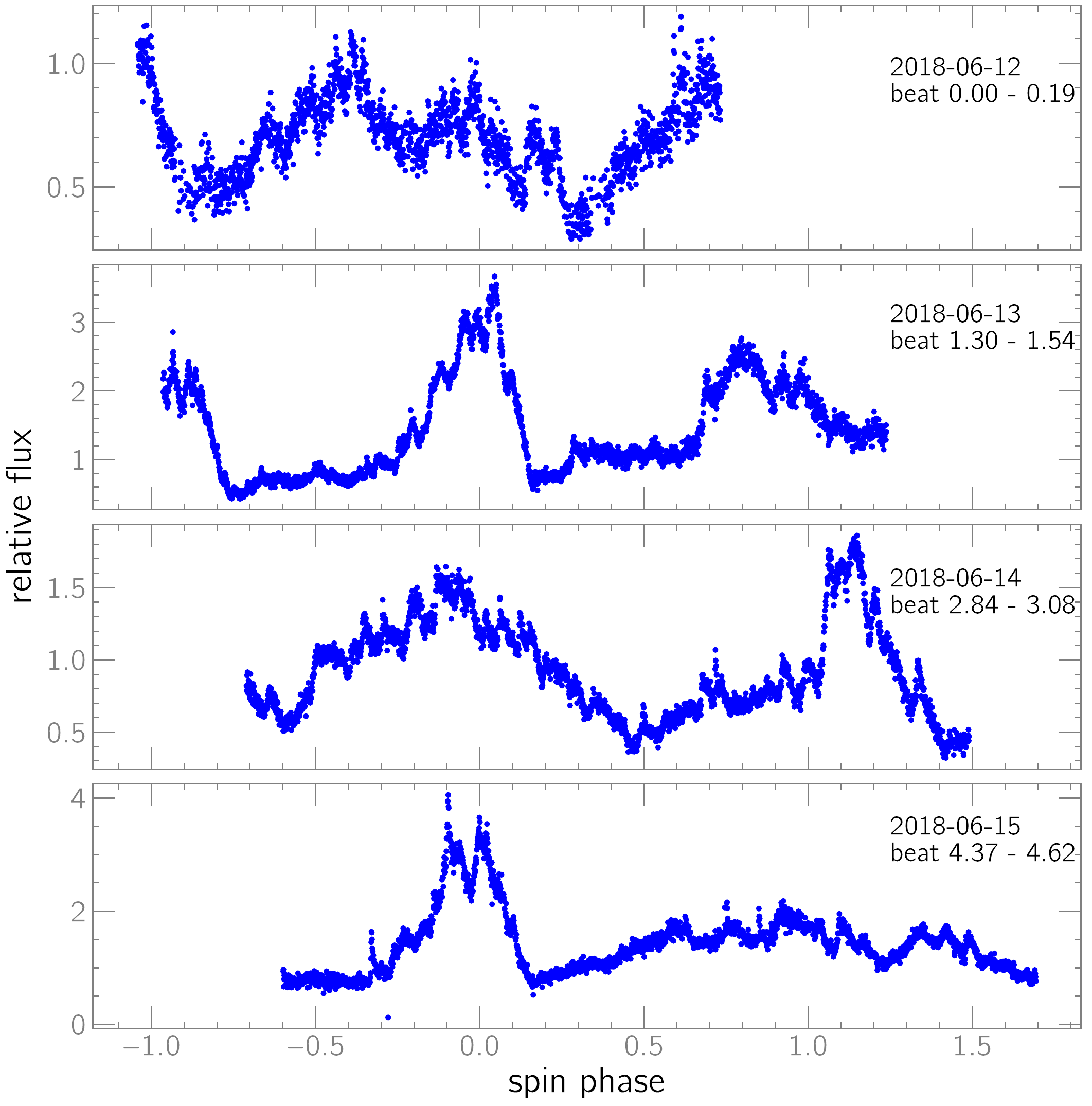}
		\caption{High-speed photometry of J1344 obtained on four consecutive nights in 2018 June. Each panel indicates the UT date at the start of the light curve and the beat cycles covered by each light curve. Spin phase 0.0 was arbitrarily chosen to correspond with the strongest peak in the light curve from 2018 June 13, while beat phase 0.0 corresponds with the first point in the June 12 light curve.
		}
		\label{fig:ground_lc}
	\end{figure*}
	
	Although APs once had very distinct \ratio\ ratios compared to IPs, the unusual systems Paloma \citep[\ratio=0.87;][]{schwarz, joshi, littlefield22} and \swift\ \citep[\ratio=0.8;][]{halpern, rawat22} have blurred the once-clear line of division between the APs and IPs.\footnote{\citet{halpern} identifies two possible sets of frequency identifications, resulting in either \ratio=0.80 or \ratio=0.89. We agree with \citet{rawat22} that the available data favor \ratio=0.8 for \swift, but in light of the complexities discussed by \citet{halpern}, \ratio\ still requires observational confirmation. } Throughout this paper, we use the neutral term ``asynchronous mCV'' to refer to systems in which it is unknown whether the observed asynchronous rotation is an equilibrium condition.\footnote{The existing terminology for Paloma-type systems is inelegant, as they are often referred to as either ``nearly synchronous IPs'' or ``highly asynchronous polars.''}
	
	Long, uninterrupted light curves of asynchronous mCVs are of great value because they provide coverage of the beat period between the spin and orbital periods. The beat period is time required for the WD to complete a single rotation within the binary rest frame, and the differential rotation modulates the geometry of the accretion flow. Because this period can be days, weeks, or even months long, it is often infeasible to observe from the ground. The Kepler spacecraft and the Transiting Exoplanet Survey Satellite (TESS) have observed several APs, resulting in a flurry of publications about the APs: CD~Ind \citep{hakala, littlefield19, mason20}, V1500~Cyg \citep{wang21}, BY~Cam \citep{mason22}, and SDSS~J0846 \citep{littlefield22}, as well as Paloma \citep{littlefield22} and \swift\ \citep{halpern, rawat22}.
	
	\subsection{\fullname}

	The subject of this study, \fullname\ (hereafter J1344), presents a challenge to these classifications. J1344 has been included in several photometric and spectroscopic studies, but it has not received in-depth attention. \citet{szkody11} reported its SDSS spectrum and classified it as a likely polar, based on a conspicuous cyclotron hump, the presence of H$\alpha$ and H$\beta$ emission, and a large ($\sim400$~km/s) radial-velocity amplitude. Follow-up time-series photometry by \citet{szkody14} found that the light curve changed dramatically over the course of days. The same study also reanalyzed the SDSS spectrum and proposed a candidate field strength of 65~MG. Most recently, \citet{thorstensen20} reported additional spectroscopy and photometry, identifying a likely orbital period of 102~min. The \citet{thorstensen20} spectrum showed He~I and He~II emission, neither of which was visible in the SDSS spectrum. 
	
	\citet{bj21} estimate J1344's distance to be $599^{+53}_{-46}$~pc using geometric priors and the Gaia eDR3 parallax \citep{gaia, edr3}.

	\section{Data}

	The Transiting Exoplanet Survey Satellite (TESS) observed J1344 from 2022 March 26 until 2022 April 22 at a 2~min cadence. The TESS pipeline produced two light curves of J1344: one that presents the simple-aperture-photometry (SAP) of J1344 and another that shows the preconditioned SAP (PDCSAP) flux. Both are based on the same underlying observations, except that PDCSAP flux attempts to correct the SAP light curve for systematic artifacts. We saw no significant differences in the two light curves and elected to use the SAP data for our analysis. We specified a ``hard'' quality bitmask to exclude sections of the light curve of suspect quality, resulting in several large gaps in the light curve (Fig.~\ref{fig:power}). 
	
	To establish the historical context for J1344's accretion rate during the TESS observations, we downloaded the ATLAS light curve. The ATLAS data (Fig.~\ref{fig:power}) show that J1344 was near magnitude 19 in both the c- and o-bands during the TESS observation. For context, J1344 was approximately one magnitude brighter in 2016 February-March, when \citet{thorstensen20} obtained time-series spectroscopy. Thus, the \citet{thorstensen20} spectra were obtained when the accretion rate was higher.
	
	Our dataset also includes four previously unpublished light curves of J1344 from 2018 June 12, 13, 14, and 15. These data were obtained with the Otto Struve (2.1-m) telescope of the McDonald Observatory at a cadence of 5~sec, without dead time. A broad-band optical filter covering the Johnson BVR range was used and the images were dark-subtracted and flat-field-corrected using routines in Python. The nightly light curves are shown in Fig.~\ref{fig:ground_lc}. 
	
	\begin{figure*}
		\centering
		\includegraphics[width=\textwidth]{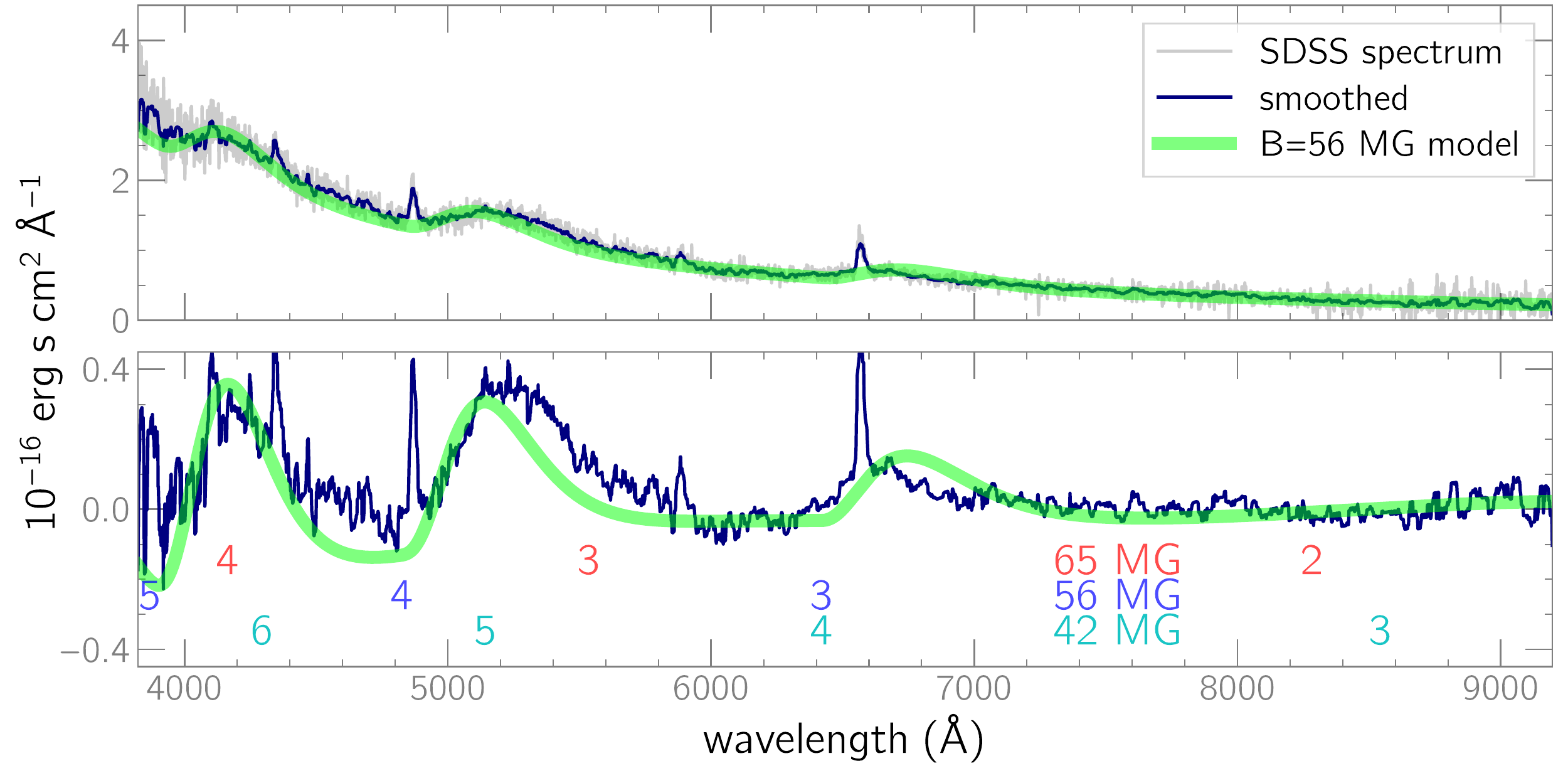}
		\caption{{\bf Top:} SDSS spectrum of J1344, with the best-fit cyclotron model spectrum. {\bf Bottom:} Median-smoothed spectrum from the top panel after subtraction of the continuum. Cyclotron harmonics near 4100\AA, 5200\AA, and 6600\AA\ are visible. The expected positions of the $n$-th cyclotron harmonics are labeled for three different field strengths: the original estimate of 65~MG (shown in red) from \citet{szkody11}, 56~MG (blue) and 42~MG (cyan). The 65~MG field cannot account for the hump near 650~nm, while a 42~MG field would produce a hump near 860~nm. The 56~MG model (green line), in contrast, successfully predicts the three humps that are visible.
		}
		\label{fig:cyclotron}
	\end{figure*}

	\begin{figure}
		\centering
		\includegraphics[width=\columnwidth]{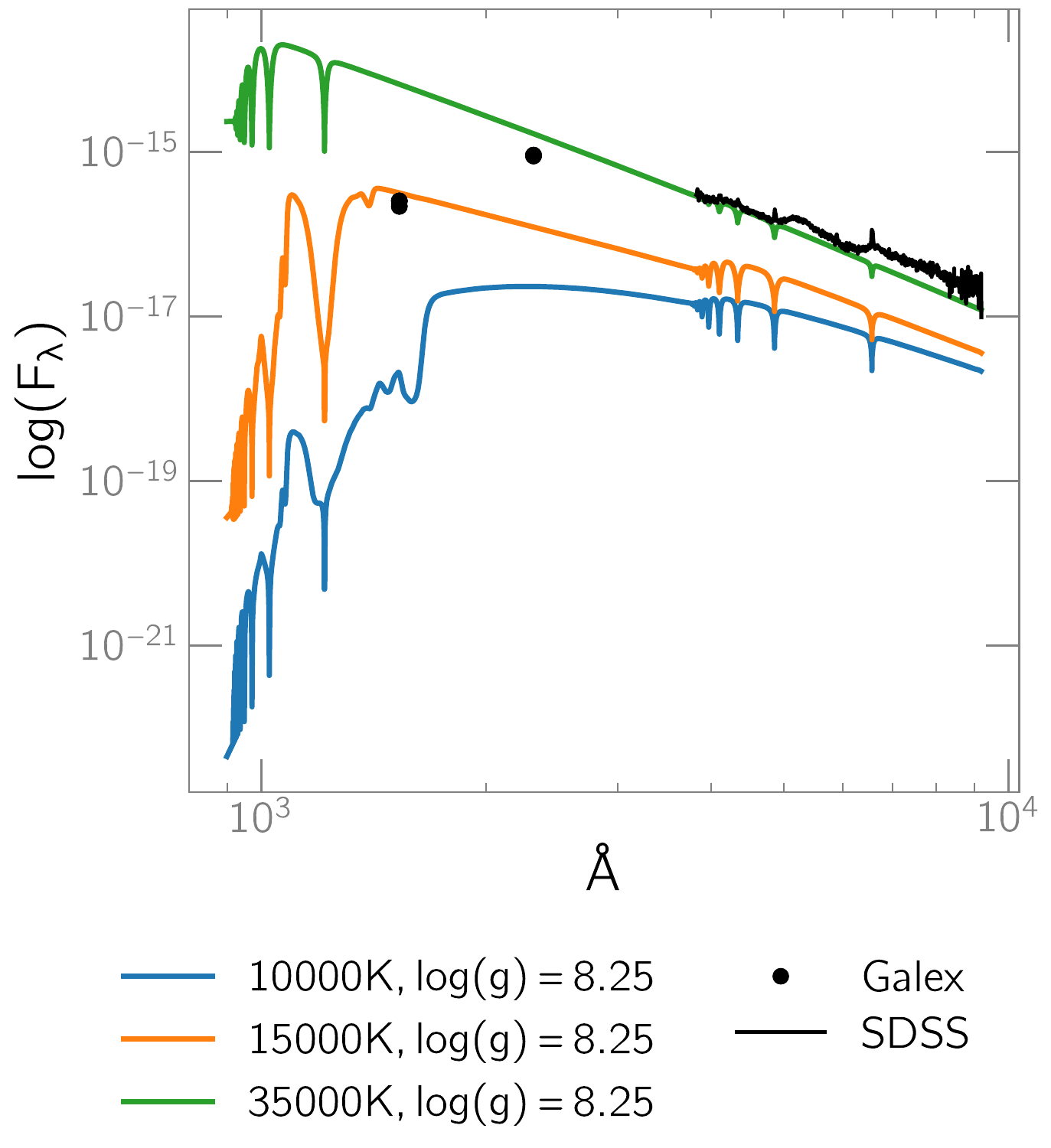}
		\caption{Spectral energy distribution of J1344 compared to selected \citet{koester} WD spectral templates for M~=$\sim0.8~M_{\odot}$, scaled to the Gaia eDR3 distance of J1344. The SED therefore cannot be explained solely by the WD's photospheric contribution; while a hot WD could explain the optical continuum, it significantly overestimates the Galex NUV/FUV measurements. Furthermore, while the SED does not pinpoint a unique WD effective temperature, it suggests that the WD is not particularly hot, providing evidence against a recent nova.
		}
		\label{fig:SED}
	\end{figure}

	\section{Analysis}
	
	\subsection{TESS photometry}
	
	Unfortunately, the signal-to-noise ratio of the TESS light curve is too low to permit an in-depth study of its temporal evolution, but the power spectrum of the full dataset (Fig.~\ref{fig:power}) contains a series of high-frequency sidebands and harmonics, similar to the TESS and K2 power spectra of other asynchronous polars. If J1344 were synchronous, we would observe only the orbital frequency $\Omega$ and its harmonics, so the presence of the sideband frequencies unequivocally establishes that $\omega \neq \Omega$.
	
	While it is relatively easy to establish the presence of asynchronous rotation, ascertaining the identities of the various signals is a more fraught process. The largest-amplitude signal is at  14.166~\cpd, and it is identical to the \citet{thorstensen20} spectroscopic period of $P_{spec} = 0.070592(4)$~d (102~min), measured from the radial-velocity variations of the H$\alpha$ emission. \citet{thorstensen20} interpreted this as the orbital period of the binary, which would be the proper interpretation in a synchronous polar (as J1344 was believed to be at the time). However, in an asynchronous system, it is possible that some (or even most) of the emitting material is trapped in the WD's magnetosphere, which would cause its radial-velocity variations to be modulated at $\omega$ rather than $\Omega$. Encountering this very problem in a study of the asynchronous mCV \swift, \citet{halpern} argued that a spectroscopic period based on the entire line will most likely yield the orbital period rather than the spin period. However, he also pointed out that this interpretation is circumstantial. 
	
	We think that the spectroscopic frequency in J1344 corresponds to the WD spin. The main justification for this inference is the high surface field strength of the WD (B$=56\pm2$~MG; see Sec.~\ref{sec:fieldstrength}) and the rather modest accretion rate during the \citet{thorstensen20} spectroscopic observations (as evidenced by the modest strength of He~II $\lambda$4686\AA\ in his spectra). A high field strength, combined with a moderately low mass-transfer rate, will reduce the extent of the ballistic region of the accretion flow (which rotates at $\Omega$) and increase the size of the magnetically confined flow (which rotates at $\omega$).

	We also considered the possibility that the signal at 14.166~\cpd\ is the $2\omega-\Omega$ sideband, as \citet{littlefield19} and \citet{mason20} argued for CD~Ind. However this identification in J1344 appears unlikely for two reasons. First, it would require that the orbital frequency and its harmonics have almost no power in the TESS power spectrum, even though $\omega$ would not be similarly impacted. Second, following the approach of \citet{mason20}, we checked whether the dominant short- and long-term periodicities in the light curve agree. We did this by computing power spectra in 0.2-d segments, summing them, and comparing the resulting power spectrum against that of the full dataset. The rationale of this exercise, as explained by \citet{mason20}, is that pole switching redistributes power from its intrinsic frequency ($\omega$) into sidebands (particularly $2\omega - \Omega$); if the light curve is divided into short segments that contain no pole switching, their power spectrum should show the true spin frequency. We find that with J1344, the short- and long-term periodicities agree, which suggests that the power spectrum in Fig.~\ref{fig:power} has not been contaminated.

	The beat frequency (\beat\ = 1.52 \cpd) is directly observable in the TESS power spectrum, and since it is defined as the difference between the spin and orbital frequencies, our tentative identification of $\omega$ enables us to identify the likely orbital frequency as $\Omega = $12.641~\cpd\ by simply subtracting \beat\ from $\omega$. The resulting \ratio\ is 0.893. If the \citet{thorstensen20} spectroscopic period is instead identical to $\Omega$, \ratio=0.903. Thus, even if we have confused $\omega$ and $\Omega$, the net result would be that J1344 is only slightly closer to being synchronous.

	\subsection{Ground-based photometry}

	The asynchronous rotation revealed by TESS can be easily reconciled with previous photometric observations and with the four light curves presented in Fig.~\ref{fig:ground_lc}. Together, Fig.~4 in \citet{szkody14} and Fig.~18 in \citet{thorstensen20} show seven light curves of J1344 from different nights, and their morphology varies profoundly. The same is true of our ground-based photometry (Fig.~\ref{fig:ground_lc}). This behavior is common for asynchronous polars \citep[e.g.][]{mason89, hakala, littlefield19} because the accretion flow couples to different field lines throughout the beat cycle in response to the asynchronous rotation. Our identification of J1344 as an asynchronous mCV therefore offers a simple interpretation of the seemingly erratic behavior noted by previous studies. Indeed, in Fig.~\ref{fig:ground_lc}, the profile of the light curve can be seen returning to its original shape after the passage of an integer number of beat cycles.
	
	In addition, the ground-based light curves confirm that the variability observed in the TESS light curve is attributable to J1344 as opposed to a blended background source.

	\subsection{Magnetic field strength} \label{sec:fieldstrength}
	
	The SDSS spectrum from \citet{szkody11} is of great value because it captured J1344 in a low-accretion state. Only five emission lines (H$\alpha$, H$\beta$, H$\gamma$, H$\delta$, and He~I~$\lambda5876$\AA) are detectable in that spectrum, and the weakness of these lines, combined with the complete absence of He~II emission, suggests that there was no accretion shock at the time of the SDSS observation. This is possible only at extremely low accretion rates, though the presence of an obvious cyclotron hump near 520~nm confirms that there was very modest accretion onto the WD when the SDSS spectrum was obtained. In contrast, the time-series spectroscopy presented in \citet{thorstensen20}, which was obtained during a state of increased accretion compared to the SDSS spectrum, shows significant He~II emission, and only the 520~nm hump is discernible.
	
	The field strength of the cyclotron-emitting region can be estimated from the wavelengths at which cyclotron humps appear. \citet{szkody11} identified two humps in the SDSS spectrum consistent with the third and fourth cyclotron harmonics produced in a 65~MG field. However, they pointed out that the n=2 harmonic near 8200~\AA\ was absent in the SDSS spectrum, even though the SDSS spectrum provides adequate wavelength coverage. Our reexamination of the SDSS spectrum favors a somewhat lower field strength. Fig.~\ref{fig:cyclotron} shows the presence of at least three cyclotron humps in the SDSS spectrum near 4200~\AA, 5200~\AA, and 6600~\AA. We find that these three humps can be explained as harmonics 3-5 in a B$\sim$56~MG field or as harmonics 4-6 at B$\sim$42~MG. 
	
	To distinguish between these two possibilities, we fit the SDSS spectrum with a homogeneous cyclotron-spectrum model \citep{cyclotron}, a technique that is more precise than simply computing the approximate wavelengths of cyclotron humps. A cyclotron spectrum was calculated for magnetic fields between 40 and 70~MG, with the wavelengths of the harmonics best matching the humps in the observed spectrum near 56~MG. A value of 55~MG better matches the n=4 harmonic peak while 57~MG improves the fit to the n=3 harmonic peak. Therefore, we estimate the uncertainty on the field to be $\pm 2$~MG. The cyclotron spectrum was added to a continuum function varying as $\lambda^{-\alpha}$, and the electron temperature $T$ was varied to approximate the width of the cyclotron humps. A temperature of $kT\approx 12\pm 3$~keV was found to provide a fair match to the widths.  The best-fit cyclotron spectrum, shown in Fig.~\ref{fig:cyclotron}, has a field strength of $B=56\pm2$~MG, and we adopt this as the surface field strength of the WD. 
	
	Does the photosphere of the WD contribute significantly to the observed SDSS spectrum? To investigate this possibility, we scaled the \citet{koester} spectral templates for non-magnetic WDs, varying both the effective temperature and surface gravity, to the Gaia distance of J1344 (599~pc). The three-dimensional extinction maps of \citet{green} indicate negligible reddening along the line of sight to J1344, so there is no need to deredden the SDSS spectrum, which we find to be significantly brighter than expected for a WD at that distance (Fig.~\ref{fig:SED}). \footnote{We neglect the spectroscopic contribution of the (presumed) mid-to-late M companion star, as it is not a plausible origin of the blue continuum and is absent even at the red end of the SDSS spectrum.} The scaled spectral templates can reach the SDSS continuum only for implausibly low WD masses or WD temperatures that are so high they they contradict the Galex NUV/FUV measurements. We conclude that the blue continuum of the SDSS spectrum cannot be attributed exclusively to the WD photosphere, which suggests that cyclotron radiation is the most likely culprit for the undulations in the SDSS spectrum, consistent with the interpretation of \citet{szkody11}.

	\section{Discussion}
	
	\subsection{Comparing J1344 to other mCVs}
	
	What makes J1344 remarkable is the combination of three parameters: its moderately high magnetic-field strength, its short orbital period, and its high degree of asynchronism. Together, these properties paint an interesting picture of a magnetic CV that would normally be expected to be synchronous, but isn't.
	
	\begin{figure}
		\centering
		\includegraphics[width=\columnwidth]{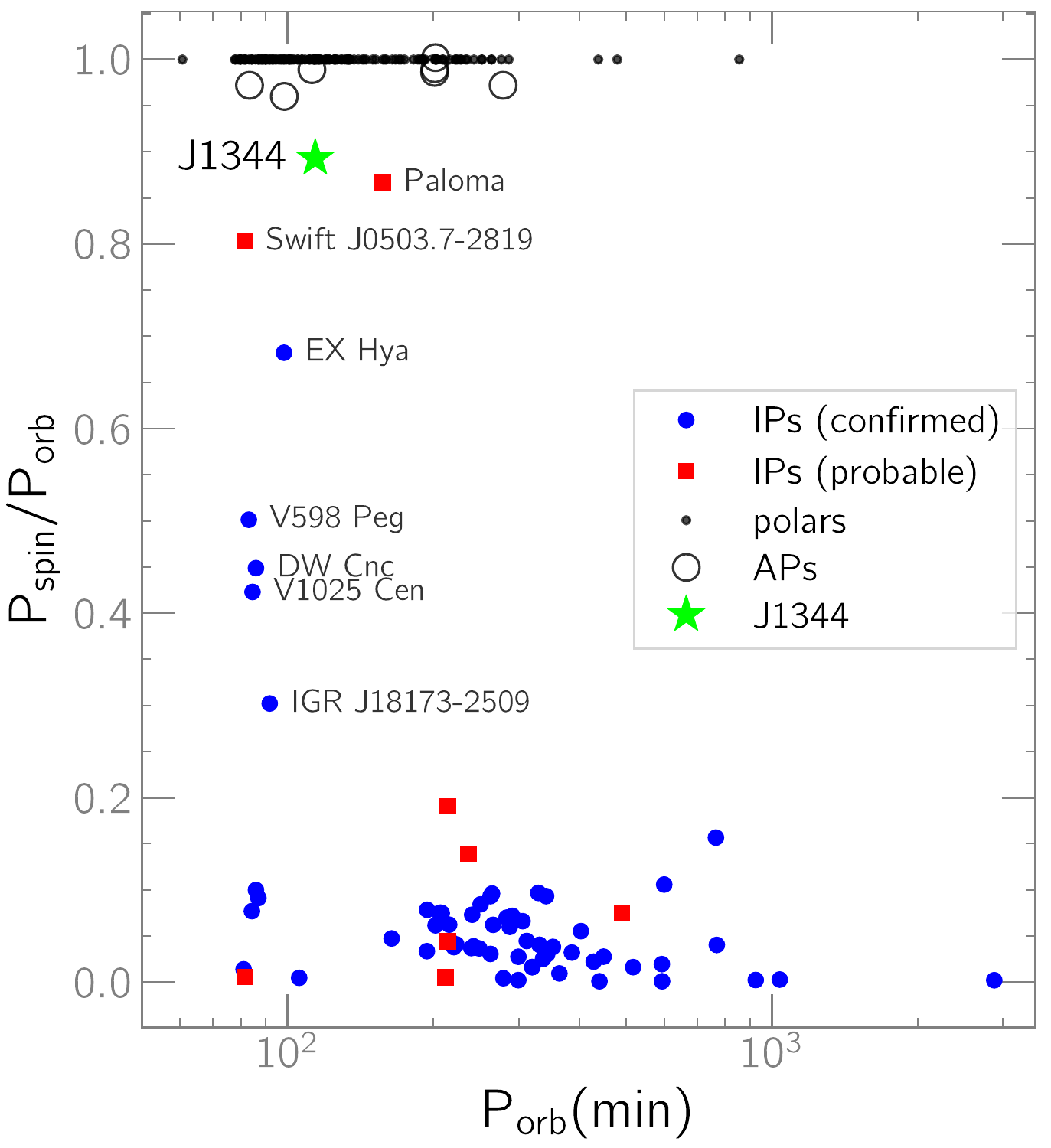}
		\caption{\ratio\ as a function of orbital period in mCVs. The IP data were downloaded from Koji Mukai's IP catalog; blue markers represent the IPs assessed by Mukai to be either ``confirmed'' and ``ironclad,'' while red markers are ``probable'' IPs. We use Mukai's compilation of spin periods, except for Swift~J0503.7-2819 and Paloma, whose spin periods are from \citet{rawat22} and \citet{littlefield22}, respectively. The AP data are from Table~1 in \citet{littlefield22}, and the periods of the polars are from the AAVSO VSX catalog. We have labeled J1344 as well as individual mCVs with $0.25\leq$\ratio$\leq0.9$.
			\label{fig:IPs}}
	\end{figure}
	
	J1344 spin-to-orbit ratio of \ratio=0.893 is most unusual for an mCV. In IPs, \ratio$\lesssim0.1$ is very common, and only a handful of short-orbital-period IPs show \ratio$>>$0.1  (Fig.~\ref{fig:IPs}). EX~Hya has long been the benchmark system in this regard, and its unusual \ratio\ of 0.68 has received significant theoretical attention \citep[e.g.,][]{ex_hya}. While EX~Hya is famous for being a slowly rotating IP in a unique spin equilibrium \citep{ex_hya, norton}, Paloma \citep[\ratio=0.87;][]{schwarz, joshi, littlefield22} has been known for nearly two decades as a solitary outlier that blurs the dividing line between IPs and APs. But with the recent identifications of J1344 and \swift\ \citep{halpern, rawat22}, it is becoming clear that Paloma's \ratio\ is not as extreme as it might have once appeared (Fig.~\ref{fig:IPs}).
	
	Likewise, the magnetic-field strength in J1344 is notably high in comparison to the relatively few asynchronous mCVs for which an estimated field strength has been published.\footnote{We do not consider the detached system AR~Sco because its field strength has not been conclusively established.} The detection of polarized emission in spectropolarimetry has led to estimates of $\sim9-27$~MG \citep{vaeth} and B=$31.5\pm0.8$~MG in V2400~Oph and V405~Aur \citep{piirola}, respectively. As for the APs, our review turns up four systems with reliable magnetic-field estimates: $11\pm2$~MG in CD~Ind \citep{schwope}, 40.8~MG in BY~Cam \citep{cropper_bycam}, B$\lesssim$20~MG in \igr\ \citep{tovmassian}, and B$_1$=72~MG, B$_2$=105~MG for two separate accretion regions in the old nova V1500~Cyg \citep{harrison18}. Indeed, J1344's field strength is more typical of synchronous polars \citep{ferrario15}. Unfortunately, the magnetic-field strength has not been measured in either Paloma or \swift, the two mCVs that have the most similar values of \ratio\ to that of J1344.

	The third distinguishing property of J1344 is that it has a short orbital period and therefore a relatively small binary separation. As the binary orbit shrinks, the secondary will interact more strongly with the WD's magnetosphere, which greatly facilitates synchronization \citep{chanmugam}; moreover, the secular mass-transfer rate below the period gap is significantly lower than above the gap \citep[e.g.,][]{knigge}, which reduces the spin-up torque caused by accretion. As we discuss in the following subsection, these properties tend to favor synchronization.

	\subsection{The nature of asynchronous rotation in J1344}
	
	As noted earlier, a short orbital period increases the synchronization torque while decreasing the opposing accretion torque, and this principle has led to speculation that IPs can evolve into polars when they cross the period gap \citep{chanmugam}. Although Fig.~\ref{fig:IPs} might seem to lend credence to this theory by showing that high-\ratio\ mCVs occur exclusively at short orbital periods, this appearance is misleading, as (1) there is a predicted continuum of rotational equilibria for EX Hya-like systems \citep{ex_hya, norton} and (2) the synchronization timescales in \citet{chanmugam} and \citet{schreiber} are several Myr, which is very short compared to the multi-Gyr lifetime of a CV. The probability of detecting a system synchronizing for the first time is therefore rather low.
	
	An important step towards understanding the asynchronous rotation in J1344 is knowing whether the WD is in rotational equilibrium or is instead evolving towards synchronous rotation. Without knowledge of the long-term behavior of J1344's spin-period derivative, we must remain agnostic as to whether the system is in rotational equilibrium, which occurs when the spin-up torque of accretion is balanced by the countervailing drag of the magnetic field through the accretion flow and has been a major focus of IP theoretical research \citep[e.g.,][]{wynn95, ex_hya, norton}. However, authoritative theoretical studies by \citet{norton, norton08} hypothesized that Paloma-like mCVs with large \ratio\ ratios contain secondaries with unusually low magnetic moments and were therefore exceptions to the studies' models of rotational equilibrium.

	\citet{norton} predict that synchronization will occur when the magnetic torque exceeds the accretion torque \citep[see also][]{chanmugam}. So,
	\begin{equation}
	\frac {\mu_{1}\mu_{2}} {a^{3}} > \dot{M} \sqrt{GM_{1}R_{m}},
	\label{eqn:sync}
	\end{equation}
	where $\mu_{1}$ and $\mu_{2}$ are the respective magnetic moments of the WD and the secondary, $a$ is the binary separation, $G$ is the gravitational constant, $M_{1}$ is the WD mass, and $R_{m}$ is the magnetospheric radius \citep[][their Eq.~13]{norton}. Within this framework, there are two scenarios under which an mCV could avoid synchronization: a large accretion torque (represented by the right side of Eq.~\ref{eqn:sync}) or a low magnetic-locking torque (described by the left side of the equation). The former appears unlikely, based on the weak or absent He~II emission in all published spectra of J1344; moreover, the system's absolute magnitude of $G=9.2$ is consistent with low-luminosity IPs (Mukai \& Pretorius, in prep.). This suggests a low mass-transfer rate, which is unsurprising in light of the well-established tendency for mass-transfer rates to be $\sim2$ orders of magnitude lower in CVs below the period gap compared to systems above the gap \citep{knigge}.
	
	A lower-than-usual magnetic-locking torque is a more promising explanation, and there are at least two mechanisms that could result in this. First, as \citet{norton, norton08} speculated, the secondary's magnetic moment $\mu_{2}$ might be very low. The other possibility is that if the WD were unusually massive, its magnetic moment ($\mu_{1}=Br^{3}$) would be lower than for a typical-mass WD of equal surface field strength because of the inverse relation between a WD's mass and its radius. Moreover, increasing the WD's mass would also increase the binary separation $a$, upon which the magnetic-locking torque has an inverse-cube dependence. Per Eqn.~\ref{eqn:sync}, the opposing accretion torque would also increase with WD mass, albeit with a weaker dependence ($\propto M_{1}^{1/2}$).
	
	As a quantitative illustration of the massive-WD hypothesis, we consider two hypothetical WD masses ($M_1$ = 0.8~M$_{\odot}$ and 1.3~ M$_{\odot}$) in a binary with an orbital period of 114~min and a secondary mass of $M_2=0.1~M_{\odot}$. Applying the \citet{nauenberg} mass-radius relationship to these two WD masses, the less-massive WD has a radius $\sim$2.5 times larger, so its magnetic moment would be higher by a factor of $2.5^3$. The binary separation $a$ would be $14\%$ smaller at $M_1$ = 0.8~M$_{\odot}$, and the locking torque ($\propto a^{-3}$) would increase by a factor of 1.56 compared to the higher-mass WD. Together, these two factors would result in a factor of $\sim24$ reduction in the locking torque at $M_1$ = 1.3~M$_{\odot}$ compared to $M_1$ = 0.8~M$_{\odot}$. Moreover, the diminished locking torque for the massive WD would be accompanied by a $\sim27\%$ increase in the spin-up torque from accretion. Collectively, these considerations suggest that the mass of the WD can play an important role in determining whether the synchronization requirement in Eq.~\ref{eqn:sync} is satisfied.
	
	Given the strong dependence of the locking torque on the WD mass, it will be important for a future study to measure or constrain the WD mass. If the WD turns out to be of typical mass, it would provide indirect support for the \citet{norton, norton08} proposal that the secondary has an unusually low magnetic moment $\mu_2$.

	It is also possible that J1344 is a formerly synchronous polar that is out of equilibrium and in the process of resynchronizing, although we disfavor this possibility for several reasons. First, the most commonly invoked method of desychronizing a polar is a nova eruption, but there is no evidence of a nova eruption in J1344. The Digital Access to a Sky-Century at Harvard collection of photographic plates \citep{grindlay} shows no detections at J1344's position, and the SED (Fig.~\ref{fig:SED}) is inconsistent with a hot WD. Moreover, the recurrence timescale for novae becomes increasingly long at short orbital periods due to the diminished mass-transfer rates at those periods \citep[e.g., Fig.~5 in ][]{knigge}, so the likelihood of observing such a system by chance is low. Another difficulty with the nova hypothesis is that the amount of angular momentum required to spin up a previously synchronized WD by a minimum\footnote{If J1344 is in a temporarily desynchronized state due to a long-ago nova eruption, we would expect it to have been even more desynchronized immediately after the nova.} of 10\% is immense, particularly when one considers that the prototypical AP, V1500~Cyg, was desynchronized by only 3\% following its nova eruption.
	
	Finally, as a corollary to the previous paragraph, it is also conceivable that J1344 is synchronizing for the first time---but here again, we encounter a difficulty with the low probability of observing such a system in a short-lived stage in its overall lifetime. \citet{schreiber} proposed that polars are the descendants of \textit{non-magnetic} CVs whose WDs began to crystallize after having been spun up by accretion; the emergence of the magnetic field causes the formerly non-magnetic CV to synchronize rapidly. Even the slowest synchronization timescales contemplated by \citet{schreiber} make it improbable that such a system would be serendipitously discovered during this brief process.

	\section{Conclusion}
	
	We have reclassified the nominally synchronous polar \fullname\ as an asynchronous mCV with \ratio=0.893, which is simultaneously much more desynchronized than observed in APs but also much more synchronized than seen in IPs. Based on its unusual \ratio, J1344 bears a striking resemblance to \swift\ (\ratio=0.8) and Paloma (\ratio=0.87), but unlike those systems, J1344's magnetic field strength is easily measurable (B=56$\pm$2~MG). The highly asynchronous nature of J1344 reveals that the combination of B$\gtrsim10$~MG and a short binary separation does not guarantee that a system will rapidly synchronize. Within the existing theoretical framework, some combination of a weakly magnetic secondary star or an unusually massive WD is the most attractive explanation for the failure of J1344 to achieve synchronous rotation.

	We have tentatively identified the \citet{thorstensen20} spectroscopic frequency as the WD spin frequency, but follow-up studies are needed to confirm this. These studies should seek out spectral features that are unambiguously attributable to the secondary (either emission lines from its inner hemisphere or absorption features), although this will be challenging given the late spectral type expected for the secondary at $P_{orb}=114$~min. Even if it turns out the \citet{thorstensen20} frequency is the orbital frequency, our results are not seriously impacted, and the spin-to-orbit ratio would increase only slightly to \ratio=0.903. 
	
	Finally, J1344's ability to successfully masquerade as a synchronous polar for a decade suggests that other nominally synchronous systems might also be asynchronous. It will be important to examine the TESS light curves of all polars to search for any such systems.

	\begin{acknowledgements}
		
		During the preparation of this manuscript, the astronomical community experienced a profound loss with the untimely death of Tom Marsh. Over the course of his career, Tom established himself as both a distinguished researcher and a cherished colleague. His legacy will be felt for many years to come.
		
		We thank Koji Mukai for helpful comments and for sharing a draft of a forthcoming paper.
		
	\end{acknowledgements}

	\facilities{TESS, Struve, Sloan}
	
	\software{ {\tt astropy} \citep{astropy18}, {\tt lightkurve} \citep{lightkurve} }

	\bibliography{bib.bib}

\end{document}